\documentstyle[prd,aps,preprint]{revtex}

\begin{document}
\draft

\preprint{Imperial/TP/94-95/39}


\title{Winding Number Correlation Functions and Cosmic String Formation}
\author{L. M. A. Bettencourt, T. S. Evans and R. J. Rivers}
\address{The Blackett Laboratory, Imperial College,
London SW7 2BZ, U.K.}
\date{\today}
\maketitle

\begin{abstract}
We develop winding number
correlation functions that allow us to assess the role of field
fluctuations on vortex formation in an Abelian gauge theory.
We compute the behavior of these
correlation functions in  simple circumstances and show how
fluctuations are important in the vicinity of the phase
transition.  We further show that, in our approximation,
the emerging population of long/infinite string
is produced by the classical dynamics of the fields alone, being essentially
unaffected by field fluctuations.

\end{abstract}



\section{Introduction}

The production of topological defects seems to be an inescapable
consequence of phase transitions in field  theories with a topologically
non-trivial vacuum manifold \cite{Kib,Book}.
This is  observed experimentally in numerous
condensed matter systems from superfluid Helium\cite{Helium}
and superconductors
to ordinary liquid crystals \cite{liqcrys} and is expected to occur
in the early universe
for most theories of Grand Unification. Defects
produced at such energy scales are strong candidates to
generate the observed large scale structure of the universe, competing
with the fluctuations of inflationary models to match the
observations\cite{Struc}.

In spite of their universality, the problem of predicting the
details of networks of defects emerging from a phase transition
is still unsolved, being  necessarily
entangled with the more general question of how to determine the dynamical
evolution of fields throughout the transition.
Our qualitative picture of defect production is given by the
well-known Kibble mechanism \cite{Kib}. With the $U(1)$ theory of a
complex scalar field in
mind, it essentially assumes that the phase
transition proceeds by the formation of domains, which are
characterized by some random value of the field phase, that later
coalesce to complete the transition. Such domains, at the time of
coalescence, can, in turn, be characterized by their average radial
length scale $\xi$, which in most applications is taken to be the
correlation length as computed for a massive scalar field in
thermal equilibrium \cite{Kib}.
As domains coalesce phase gradients must be minimized and integer
windings can be formed. Because of its simplicity this picture has great
appeal and is amenable to a straight-forward approximate
numerical implementation
by the so-called Vachaspati-Vilenkin algorithm \cite{VV}, which in
turn constitutes  the usual starting point for numerical network
evolutions.

As it stands the Kibble mechanism is a wholly static argument,
failing to answer any questions
regarding the role of fluctuations in the fields. If important enough,
such fluctuations will necessarily alter the number and distribution
of defects produced by the dynamics of domain coalescence and could
thus distort seriously the predictions solely based upon it.
A change in initial string density and especially in the amount of
long string versus that in small closed loops, for example,
necessarily affects the
transient regime that characterizes a string network prior to scaling.
Moreover, it
is precisely the point at which thermal fluctuations cease to be able
to change string configurations appreciably that determines the emerging
 defect network. This problem has been attracting great enthusiasm
leading to a number of different approaches \cite{KibVil,Brand,Riv}.

Our aim in this paper is to calculate the effect of fluctuations on
winding number, requiring us to
develop
 quantities and criteria for describing fluctuations about general
field configuration solutions to the phase transition dynamics.
We focus our analysis on cosmic strings in the Abelian Higgs model and
consider correlation functions of the winding number (defined as the number
of quantized units of magnetic flux) in several circumstances.
Even
though the present study is motivated by cosmological questions the
analogy with superconducting systems is almost complete, because of
the role of the gauge fields in both models. As such
our results should apply equally well there,  where contact
with experiment can prove a good test on the ideas developed herein.
In principle the interpolation between both pictures can be obtained
essentially by
replacing cosmic strings with vortex lines and discarding some specific
aspects of relativistic self-energies in section 3.  In practice we
have yet to so so.

The outline of the paper is as follows. We start, in section 2,
by describing the
qualitative aspects of the classical dynamics of the  fields during
the phase transition and determine under which conditions the
measurement of
quantized units of magnetic flux becomes a reliable estimate of the
number of strings present. We then proceed, in section 3, to compute the
fluctuations
in winding number in the simplest scenario, namely that of a
constant background scalar field with an equilibrium distribution and
under the assumption that the gauge field, too, displays a thermal
spectrum. In section 4 we show that the computations of
section 3 can  easily be extended to the calculation of  correlation
functions of the winding number threading through two different
contours with
relative specific geometries. We use these correlation functions to answer
questions concerning the conditional probability of finding string
crossing a given surface in space, subject to the condition it has crossed
another, with a given separation and orientation relative to the
former. In section 5, we return to the computation of winding number
fluctuations, by including inhomogeneities in the scalar field. We
show how a systematic expansion around a homogeneous background can be
performed, generating as a zeroth order term the result of section 3,
and compute the first order correction resulting from the presence of
string. By comparing their relative magnitude we can formulate a
well defined criteria for the breakdown of such an expansion.
In section 6 we present our conclusions and compare our results with
those of other recent approaches to the same problem.

\section{Phase transition field dynamics and defects}

We begin with the semiclassical analysis of the
simplest theory that admits gauged strings, the
Abelian Higgs model. We consider it in the symmetry broken phase so
that the decomposition of the complex scalar field into modulus
$\varphi$ and phase $\alpha$ is possible as
\begin{equation}
\phi(x) = \varphi(x) e^{- i \alpha(x)}.
\end{equation}
for non-zero $\varphi(x)$.
Thus the classical action, including the gauge fixing term, can be written as

\begin{eqnarray}
S[\varphi,\alpha, A_\mu] &=&
\int d^4 x \left[-{1 \over 4} F_{\mu \nu} F^{\mu
\nu} + {1 \over 2} \left[ \partial_\mu \varphi \partial^{\mu} \varphi
+ \varphi ^2 (e A_\mu - \partial_\mu \alpha)(e A^\mu -
\partial^\mu \alpha)\right]\right. \nonumber \\  &-& \left.
{\lambda \over 8} (\varphi^2 - \eta^2)^2 \right]
+ {1 \over 2 a}
\partial_\mu A^\mu \partial_\nu A^\nu.
\end{eqnarray}
Our objective is to compute string formation in gauge theories,
as a dynamical process occurring as a by-product
of the symmetry breaking in the theory, by
measuring units of the quantized magnetic flux crossing a surface
bounded by the contour $\gamma$,  given by
\begin{equation}
 \langle N_\gamma\rangle = {e \over 2 \pi} \int_\gamma dl^i \langle A_i\rangle.
\end{equation}

The idea of using the magnetic flux as an indicator of the net winding
number springs from the vacuum field configurations in the broken
phase, namely,
\begin{eqnarray}
\langle\varphi\rangle &=&\eta  \\
\langle A_\mu - {\partial_\mu \alpha \over e }\rangle & \equiv &
\langle A_\mu^\prime\rangle
=0.
\end{eqnarray}
This guarantees that the gauge
field indeed traces the gradient in phase of the scalar field. In a
situation where the scalar modulus is constant, however, this becomes
simply a gauge transformation, corresponding to the residual freedom
in defining $A_\mu$ in the Lorentz gauge ($\partial_\mu A^\mu =0$).
However, it is the case of a non-constant scalar modulus that is relevant
for our purposes, corresponding to the presence of strings. In this
 case the phase can change around a closed contour in space by a
non-zero multiple of $2 \pi$, implying the existence of
a singularity in its
gradient as the contour is shrunk to a point (at which there is a
necessary restoration of symmetry). The gauge field at such a point, the
core of the string, exhibits therefore its expectation value in the
symmetric phase, i.e., $\langle A_\mu\rangle=0$. The approximate solution of
the static
Euler-Lagrange equation interpolating between these two regimes is
well-known to be \cite{NO}

\begin{equation}
A_\theta = {n \over e r } -n \varphi K_1(e \varphi r),
\end{equation}

\noindent for a homogeneous scalar modulus $\varphi$. At large
distances from the the core of the string ( $r < {1 \over e \varphi}$
), the modified Bessel function $K_1$ becomes
exponentially decreasing and the vacuum of the spontaneously broken
theory is rapidly approached. In this limit, applying the form (6) in
(3) gives precisely $n$ units of flux with exponentially small
corrections.

As the phase transition proceeds and the fields become smooth we
therefore expect $A_\mu$ to trace the
overall differences in phase left over from the dynamics of the scalar
field with increasing accuracy. To see in more qualitative detail how this
happens we need to
consider the actual Euler-Lagrange equations of the theory

\begin{equation}
\partial_\mu \partial^{\mu}\varphi = \varphi \left[ (e A_\mu -
\partial_\mu  \alpha)(e A^\mu - \partial^\mu \alpha) -  {\lambda \over
2} (\varphi^2 - \eta^2) \right],
\end{equation}

\begin{equation}
\partial^\mu \left( F_{\mu \nu} + {1 \over a} \partial_\nu
A_\mu\right) = J_\nu ,
\end{equation}

\noindent and
\begin{equation}
\partial^\mu J_\mu =0,
\end{equation}

\noindent where

\begin{equation}
J_\mu =  e\varphi^2 ( e A_\mu -\partial_\mu \alpha).
\end{equation}
The phase transition is triggered by an instability in the long
wave-length modes of the scalar field\footnote{The details of the
triggering mechanism depend crucially on the hierarchy of the
couplings in the theory determining thereby  the order of the
transition, as seen from thermodynamic effective potential
constructions.}, whose amplitudes then start
growing exponentially towards $\eta$.
This is expected to happen independently in regions of
space-time which are causally disconnected for the time scales
involved. As a result a domain structure is expected to form, leading to an
inhomogeneous scalar modulus on a scale of several domains. These
inhomogeneous field configurations do not minimize energy  and
will therefore evolve in time so as to coarsen and smooth out
\footnote{This is
explicitly observed in condensed matter systems,
such as specific types of liquid crystals, which display a large
relaxation time \cite{liqcrys}.}.

What is the behavior of the phase and vector potential during this
dynamical process? The vector potential is characterized initially, in
the symmetric phase, by $\langle A_\mu\rangle=0$. During the transition $A_\mu$
is
sensitive to the change in expectation values of $\varphi$ and $\alpha$
in quite a different manner. Whereas the change in the expectation
value of $\varphi$ gives a mass to $A_\mu$, without changing its
expectation value, the presence of a gradient in phase does act as a
source for the latter. This can be seen by writing (8) as

\begin{equation}
[\partial^\mu \partial_\mu A_\nu -(1-{1 \over a}) \partial_\nu
\partial^\mu A_\mu] - e^2 \phi^2 A_\nu = J_\nu^{\rm ext} - e \varphi^2
\partial_\nu \alpha
\equiv J^\prime_\nu,
\end{equation}
where $J_\nu^{\rm ext}$ is any external source, imposed on the system.

The general  $\langle A_\mu\rangle$ can, in principle, be computed
from the generating functional
for the theory which in turn follows from expanding the action (2)
around a field configuration that satisfies the
Euler-Lagrange equations with boundary conditions characterizing the
fields before and after the phase transition, as discussed above.

By introducing (11) in the action, we obtain

\begin{eqnarray}
S[\varphi,\alpha, A_\mu] &=& - {1 \over 2} \int
d^4 x  \left[ \left( e \varphi^2 \partial_\mu \alpha -
J_\mu^{\rm ext} \right)
A^\mu + {\rm terms \ in \ \varphi \ and \ \alpha} \right] \nonumber \\
 & \equiv &
 {1 \over 2} \int d^4 x  \left[
{J^\prime_\mu}
A^\mu + {\rm terms \ in \ \varphi \ and \ \alpha} \right].
\end{eqnarray}
This can in turn be re-written exclusively in terms of the currents, as
usual, since

\begin{equation}
A_\mu(x) = \int d^4 y\,\, G_{\mu \nu}(x,y,\varphi) J^{'\nu}(y),
\end{equation}
\noindent where
\begin{equation}
\left[ \partial^\mu \partial_\mu  -(1-{1 \over a}) \partial_\nu
\partial^\mu   - e^2 \varphi^2(x) \right] G_{\mu \nu}(x,y,\varphi)
= -\delta_\nu(x-y).
\end{equation}
In the absence of external sources $A_\mu$  then becomes
\begin{equation}
A_\mu(x) = \int d^4 y\,\, G_{\mu \nu}(x,y,\varphi) e \varphi(y) \partial^\nu
\alpha (y).
\end{equation}
As the phase transition approaches completion the scalar field should
become smoother, both in modulus and phase, leading to small
derivative terms relative to the mass scale. This allows us to perform
a small derivative expansion of the gauge propagator, leading to the
result
\begin{equation}
A_\mu(x) \simeq  \left( {\varphi(x)^2 \over \eta^2} \right)
{ \partial_\mu \alpha(x) \over e} + {\rm   derivative \  terms \ in \ }
 \varphi {\rm \ and \ } \partial_\mu \alpha .
\end{equation}
In the case of a homogeneous  and static
scalar modulus, this reduces to the vacuum in the spontaneously
broken phase (5).
Unlike its higher derivative corrections, the first term in (16) is
gauge independent.

Having understood qualitatively from this semi-classical analysis
 how the gauge field reacts to a
difference in phase, it is important to discuss how a gradient in phase
itself can occur and how it will evolve. Departing from the  symmetric
phase, where $\langle A_\mu\rangle =0$, eq.(9) can  be written as
\begin{equation}
\partial_\mu \partial^\mu \alpha + \left( \partial_\mu \ln \varphi ^2 \right)
\partial^\mu \alpha \simeq 0
\end{equation}
Eq (17) is homogeneous in $\partial^\mu \alpha$ showing that a gradient
in phase must exist initially for the phase to change in
space-time. Simultaneously, given that $\langle A_\mu\rangle =0$, the energy of
a
given configuration is minimized for a constant value of $\alpha$. We
therefore expect the field phase to be approximately constant
inside each domain formed by the dynamics of the
scalar modulus. Differences in
phase can then only occur when different domains come into causal
contact, and then should evolve so as to be minimized. This picture
corresponds
to the well known Kibble mechanism, in a gauge theory\footnote{ We
see that the geodesic rule comes about naturally from dynamical
considerations. Ambiguities \cite{Sriv} result from considering
the {\em static}
gauge theory in its spontaneously broken phase, where the phase
decouples and loses physical meaning.}. In  particular, according to
(17) and under  plane wave
ans\"atze for the fields, the small frequency modes of $\alpha$ will
be resonant with those of $\varphi$. As a result, given an exponential
growth in the amplitudes of $\varphi$ a corresponding exponential
decrease in $\alpha$ will occur.
As domains coalesce and $\partial_\mu \varphi$ tends to zero the remaining
 gradients in phase are frozen in and the gauge
field catches up with them, in the sense of (16).

Having this qualitative picture in mind for the description of
the field dynamics
during the phase transition we proceed, in the next sections, to compute
the effect of fluctuations on the total winding number using
the quantized magnetic
flux (3) as a probe. These are given by
\begin{equation}
\langle N_\gamma N_{\gamma^\prime}\rangle = \int_\gamma \int_{\gamma^\prime}
dl^i
dl^{\prime_j} \langle A_i(x) A_j(y)\rangle.
\end{equation}

In most of what follows we will be concerned solely with thermal
fluctuations.  Quantum fluctuations can be readily included but are
known to be subleading. Their effect is to change the value of the
vacuum expectation value of the $\varphi$-field as well as the couplings in
the theory. We will in general adopt the position of treating
$\varphi$ as providing a classical temperature dependent background field
$\langle \varphi \rangle$ in which the fields can be quantized.
In general this background will include strings.
We then need to know the form of the two point function for the gauge
fields, which in the absence of external sources, is given by
\begin{equation}
\langle T A_\mu(x) A_\nu(y)\rangle = G_{\mu \nu}(x,y,\varphi) +
\langle J^{\rho}(z)G_{\rho \mu}
(x,z,\varphi)J^{\sigma}(z^\prime) G_{\sigma \nu}(z^\prime,y,\varphi)\rangle
\end{equation}

\noindent where $G_{\mu \nu}$ is given by (14), and $J^\mu(x)=e \varphi(x)
\partial^\mu \alpha(x)$. If we assume that the contours do
not run over any strings\footnote{ This seems reasonable if the string
density is not too high. See section 5.} the
dominant term will be
\begin{equation}
\langle T A_\mu(x) A_\nu(y)\rangle = G_{\mu \nu}(x,y,\varphi)
\end{equation}
We stress that we
are not just quantizing $A_{\mu}$.  Scalar-field radiative
corrections about $\langle\varphi\rangle$ will be taken into account
in the calculation of $G_{\mu\nu}$.
The remaining uncertainty in defining the gauge propagator
explicitly results from the
unknown form of $\varphi(x)$. Having (16) in mind we will treat the
situation of a quasi-homogeneous background, expanding the propagator
in the
perturbations. The time dependence of $\varphi$ remains a problem,
however. We deal with it approximately by considering the fields in
equilibrium and changing the corresponding temperature. In the next
section we treat the simplest case, i.e., that of a homogeneous
background where the fields display equilibrium distributions.

\section{Fluctuations in Winding Number in Thermal Equilibrium}

In this section we specialize the previous discussion to the case of a
homogeneous  scalar modulus background and assume the fields to
have equilibrium distributions. The first of these conditions will be
lifted later, in section 5.
Dropping the condition of equilibrium, however, would require the
knowledge of the time dependence of the fields, in closed form, which
is unknown in the necessary detail\footnote{More is known for the
theory of a simple scalar field numerically, e.g.\cite{Boya},
but this proves of little use in our present context.}.

The degree of approximation actually involved in adopting a thermal
distribution
for the fields will necessarily depend on the dynamics of the
phase transition, which in turn will be determined by the hierarchy
of couplings in the theory. Firstly, for a theory with strong gauge couplings
 (Type I) one expects the system to depart
significantly from equilibrium (by bubble nucleation) and our approximation
will probably
grossly misrepresent the actual fluctuations in the system. In the
opposite limit, however, (strong Type II) it should hold as a good
approximation more or
less throughout the whole dynamics. The latter is the case of most high
temperature superconducting systems.
Secondly, in any case, the late stages of the transition should be
characterized by the approach to a thermal distribution and
consequently the estimates below  should describe the actual
fluctuations in the system increasingly more accurately.
This expectation is also consistent with the indications
of when the magnetic flux becomes a good indicator of the number of
strings crossing a given surface, as seen in the last section.

A strictly homogeneous scalar field is of course a necessary and
sufficient condition for the existence of {\em no} strings. It then
becomes necessary to be able to estimate how the values of
the fluctuations  computed below relate to those  in more
realistic inhomogeneous backgrounds. This issue will be addressed in
section 5, where we will show that under well-defined circumstances the
fluctuations computed here are the first term in a systematic
expansion about a homogeneous background and constitute, in general,
the dominant contribution.

In order to compute the thermal fluctuations in winding number, given
by (18), we need to know the form of the time-ordered thermal correlation
function of the $A_\mu$.
There are only two
symmetric tensors in the spatial indices which can be made in the covariant
and $R_\zeta$ gauges,
namely $\delta^{ij}, k^{i}k^j$.
The Ward or BRS identities and the Lorentz structure at finite temperature
then tell us that the most general form in covariant and $R_\zeta$ gauges is
\begin{eqnarray}
iG^{ij}(t- {t^\prime},\vec x-\vec {x^\prime}) &=&
\langle T A^i(t,\vec x) A^j({t^\prime},\vec {x^\prime}) \rangle
\\
&=& \frac{i}{8 \pi^3 \beta} \sum_{n=-\infty}^{+\infty} \int d^3 \vec k \;
e^{-i\{k_0(t-{t^\prime}) -\vec{k}.(\vec x-\vec {x^\prime}) \}} iG^{ij}(k)
\\
-G^{ij}(k) &=& \frac{1}{k^2-e^2\eta^2-\Pi_T} (\delta^{ij}
-\frac{k^{i}k^j}{k^2})
\nonumber \\ && +
\left( \frac{1}{k^2-e^2\eta^2-\Pi_L} + f(k,M,\zeta)
\right) \frac{k^{i}k^j}{k^2}
\end{eqnarray}
where the energy $k_0=2\pi i n / \beta$.
The $f$ function varies with the gauge chosen while the other
$\Pi_T$ and $\Pi_L$ terms correspond to the physical modes of the
photon, two transverse (magnetic) and a single
longitudinal (electric) mode.

{}From (18) and (21-23), taking the integrals over circular paths of radius
$L$, we see that we require two sorts of integration
\begin{equation}
\oint_L dx_i e^{-i \vec k.\vec x} , \; \; \oint_L dx_i e^{-i \vec
k.\vec x} k_i.
\end{equation}
Since the latter is zero we only pick up the term in the
propagator with the $\delta_{ij}$.  This is not surprising as we are
really looking at the magnetic field correlations and the term with
$\delta_{ij}$ is associated with magnetic fields, its self-energy,
$\Pi_T$ contains information about the magnetic screening.
Thus we have that
\begin{eqnarray}
\langle N_\gamma^2 \rangle &=&
\frac{e^2}{4\pi^2 }
\frac{i}{8 \pi^3 \beta} \sum_{n=-\infty}^{+\infty} \int d^3 \vec k \;
\frac{-i}{k^2-e^2\eta^2-\Pi_T}
\left| \oint_L d\vec{x}_i e^{-i \vec k.\vec x } \right|^2 n(\omega)
\end{eqnarray}
Note that this form is independent of the gauge chosen and the only
source of gauge dependence is in the expression used for $\Pi_T$.
Proceeding with (25), we use the leading term in the high
temperature expansion of the one-loop self-energy.  In this we are
working to leading order in the resummation scheme of
Braaten and
Pisarski \cite{BP} which is necessary at high temperatures.
Performing the energy sum leads to two terms.  There is
a pole contribution which corresponds to propagation of a
`plasmon'.   The pole is found at $k_0=\pm \omega(k)$ where
\begin{eqnarray}
0 &=&
\omega(k)^2 -e^2 \eta(T)^2 - \Pi_T(\omega(k),k)
\\
\Pi_T(k_0,k) &=& \frac{3M^2_P}{2}
\left( \frac{k_0^2}{k^2} +
\left( 1-\frac{k_0^2}{k^2} \right) \frac{k_0}{2 k}
\ln \left[ \frac{k_0-k}{k_0+k} \right] \right)
\nonumber
\end{eqnarray}
where we have worked with respect to the vacuum
\begin{equation}
\left| \frac{\partial V_{eff}}{\partial \phi} \right|_{\phi=\eta(T)}
=0   .
\nonumber
\end{equation}
The usual $O(e^2)$ result is $\eta(T)^2
= \eta^2 \left(1-{T^2 \over T_c^2}\right)$,
which vanishes at the critical temperature
$T_c$.
For abelian theories, the value of $M_p$ is
\begin{equation}
M^2_P= \frac{e^2  T^2}{3},
\nonumber
\end{equation}
for a U(1) theory with a single charged scalar or fermion. While the
position of the pole for general three-momentum is given by a
non-trivial dispersion relation, to within about 10\% this is
approximately expressible as
$ k_0^2=k^2+m^2$ where
\begin{equation}
m^2=e^2 \eta(T)^2 + M_P^2  .
\end{equation}
The difference in the final results, between choosing
this approximate dispersion relation or the true one, has
been studied numerically in some of the cases below and is found to
be negligible (less than 1\% ).

The real complication is encountered with the  cut term, running
across the zero energy point, seen in the imaginary part of $\Pi_T$
for $|k_0| < k$.  This corresponds to the physical
process of Landau damping allowed in a heat bath.  This must be
treated numerically as in this case we are not able to use any sum
rules of \cite{Braaten}, which can be seen in the logarithmic dependence on
temperature.  Equivalently, trying to look at possible $O(T)$ terms by
including only the $n=0$ term in the energy sum (e.g. see appendix of
\cite{Cop}) leads to divergent integrals and to the breakdown of the
calculation.

Doing the energy sum gives
\begin{eqnarray}
\langle N _\gamma^2 \rangle &=&
\frac{e^2 }{32\pi^5 }
\int d^3 \vec k \;
\left| \oint_L d\vec{x}_i e^{-i \vec k.\vec x} \right|^2
\left( \frac{n(\omega)}{\omega} + (\mbox{cut terms}) \right) \nonumber \\
&+& (\mbox{T=0 terms}).
\end{eqnarray}
On computing the integral in (25) we find
\begin{equation}
\langle N_\gamma^2 \rangle = {e^2 \over 4 \pi^2}L^2 \int k_\|
dk_\| d k_z {J_1^2(k_\| L) } \left( \frac{n(\omega)}{\omega} +
(\mbox{cut terms}) \right) + (\mbox{T=0 terms}),
\end{equation}
\noindent where $k_\|$ is the momentum component on the plane defined
by the contours and $J_1$ is the usual Bessel function of order 1.
This last integral cannot be fully computed analytically. For large
enough loops ($k_\| L < 1$)  we can take the large argument form of the
Bessel function.  Looking only at the pole contribution and using the
good approximation that $\omega^2=k^2+M_P^2$ leads to
\begin{equation}
\langle N_\gamma^2\rangle \simeq {e^2 \over 4  \pi^2} {2 LT \over 3}
\ln  \left({T \over m}\right) .
\end{equation}
This estimate is indeed confirmed by computing the integral (31), for
the pole term,
numerically. Figures 1 and 2 show the result of a numerical
integration looking at just the pole term with the approximate
dispersion relation $\omega^2=k^2 + m^2$  on the dependence of
$\langle N_\gamma^2\rangle$ on $LT$ and $M \over T$, respectively. The $1 \over
3$
coefficient results from the best-fit, given the functional form (32).
Using the exact dispersion relation for the pole contribution  makes
only a very slight difference to the results.  The cut term shows a
similar sensitivity on $\eta(T)$ which acts as an  infra-red cutoff.
For $L \eta(T) < 1$ the cut contribution to $N^2_\gamma$ rises as
 $L \ln L$.  However, for $L \eta(T) \sim 1$ and larger,
 a similar qualitative behavior to (32) is found, with magnetic
screening mass
\footnote{The magnetic screening mass
for abelian theories only has a contribution from
the scalar field shift.}
$m_{\rm mag} =e \eta(T)$ replacing the plasmon mass in the formula above
to give
\begin{equation}
\langle N_\gamma^2\rangle\,\, \propto\,\, {e^2 \over 4 \pi^2}
LT \ln  \left({T \over \eta(T)
}\right).
\nonumber
\end{equation}
with a coefficient of proportionality
comparable (typically, a few times larger) to that of the pole in
(32). Thus we see that the contributions from the pole and cut terms
essentially generate the same form for the winding number
fluctuations away from the critical temperature, when $\eta(T)\ll M_P$.
As such the influence of the cut term, in these circumstances,
is to change the overall coefficient in the functional form (32).
In most of what follows we will therefore restrict ourselves to
computing pole contributions.

Qualitatively the results above can be thought of as follows.
The dependence of $\langle N_\gamma^2\rangle$ on $L$ can be understood in terms
of a
random walk in the field fluctuations along the contour
perimeter.  This is commensurate with the Kibble mechanism of domain
formation, with random phases correlated on a scale $\xi = 1/e^{2}T$
(up to logarithms).

The quasi-linear dependence on the temperature ensures that, as the
system cools, fluctuations in winding number become less and less
relevant. The logarithmic dependence on the field thermal mass, in turn,
guarantees that as one approaches the critical temperature and the
mean field mass $\eta(T)$
vanishes, the magnitude of winding number fluctuations diverges. This is
certainly the instance when thermal fluctuations can
significantly change any underlying expected winding number. The
actual divergence of $\langle N_\gamma^2\rangle$ at the critical temperature
for
the pole term,
however, is an artefact of our best fit since, as can be seen in fig.
2, as the mass vanishes (for $M_P\simeq0$) the fluctuations in winding
number remain finite even though maximal. This is not the case for the
cut contribution, as far as we could find numerically where the
logarithmic behavior persists, with extraordinary accuracy, down to
$m_{\rm mag} = 10^{-4}$.

By themselves, however, the magnitude of fluctuations tell us little
about how string configurations can be changed.
This then takes us to the crucial issue  of comparing the magnitude of
fluctuations to the underlying expected winding number.
The magnitude of relative fluctuations is given by

\begin{equation}
{(\Delta N)^{2}  \over \langle \bar{N} \rangle^{2}}  =
{\langle N^2\rangle \over  \langle \bar{N} \rangle^2} -1
\end{equation}
where $\langle \bar{N} \rangle$ is the total string number, without
distinguishing strings from antistrings.
In principle $\langle N^2\rangle$ and $\langle\bar{N}\rangle^2$ must
be computed self-consistently
given a specific field background.
To estimate the behavior of (34) in our specific setting, however,
we can take our values for $\langle N^2\rangle$ and borrow the value
for $\langle\bar{N}\rangle^2$,
obtained using the Vachaspati-Vilenkin algorithm,
for the number of strings crossing a disc of area $\pi L^2$.
As required, this algorithm does not distinguish between strings and
anti-strings.
Using this value in
(34) should clearly yield an underestimate for the value of the
relative fluctuations. The expected value of the winding number thus
computed crossing a disc of radius $L$ is \cite{VV}
\begin{equation}
\langle\bar{N} \rangle = {1 \over 4} {L^2 \over \xi^2}
\end{equation}
\noindent where $\xi$ is the average domain radius at coalescence.
This can be written as an inverse of a mass scale $\xi = {1 \over
m_\xi}$, whence
\begin{equation}
{(\Delta N)^{2}  \over \langle \bar{N} \rangle^{2}} \simeq  {8 e^2 \over 3
\pi^2}
{T \over m_\xi} {1 \over m_\xi L} \ln ({T \over m}) -1.
\end{equation}
The identification of this scale with the mean domain radius at
coalescence springs directly from its numerical implementation.
Its computation
from the field theory, however,  implies a knowledge of the domain
dynamics which is, in general, very poor
\cite{Mazenko}. The original proposal by Kibble \cite{Kib} was to
identify this scale with the inverse Ginzburg temperature which is a
well defined quantity in a field theory undergoing a second order phase
transition.
More recently, Kibble and Vilenkin \cite{KibVil} argued that a
more suitable scale could be found by invoking the scaling of
the  string network emerging from the phase transition.
They discuss carefully the different length scales in the problem and
essentially conclude that
individual strings then can be identified once this typical string
separation scale becomes larger than the Ginzburg length.
Before then fluctuations in the fields are argued to be very large and
coherent long-lived field configurations cannot persist. Looking at
(36) we can indeed observe that fluctuations become large for
temperatures larger than $m_\xi$ and length scales smaller than its
inverse. At the critical temperature the relative fluctuations should
diverge due to the cut contribution discussed above.

Our present analysis thus lends support to several of their qualitative
arguments and clarifies the emergence of some scales from field theory
propagators.

\section{String spatial distribution and other correlation functions}

One interesting question to ask about string formation
that is of the upmost importance for the subsequent evolution of a
string network is what is the fraction of string in loops as opposed
to long string. We will see in this section that winding number
correlation functions can give us some insights into
this problem.

Since we expect to be dealing with a complicated background in which
many string configurations will thread
 through our contour  these
questions can only be answered statistically.
Intuitively then, the question of how string is distributed in space
assumes the form of a conditional probability question, i.e, knowing
the amount of string that crosses a given surface what is the amount
of string that crosses a second surface with a given orientation
relative to the former. Answers to these questions are naturally
provided
in terms of correlation functions and the natural
way ahead, in the context of our present discussion, is to take our
two point function for the winding number (18) through two different
contours. In  so doing we expect that the correlations
between net winding number threading through the two loops in a given
geometrical configuration can also give us information about how
string is distributed relative to that geometry.  The general case
is too difficult, but there are two natural extensions of (25) that
are solvable,
namely that for which the two circles remain concentric but now have
different radii $l$ and $L >l$ and that of two co-axial circles of equal radius
$L$ separated by a distance $h$.
These are shown in Figs. 3 and 4, respectively. Intuitively we would
expect the first of these correlation functions to give us information
about the population  of string with curvature radius greater than
that of the longest contour whereas the second should provide a
measure of straightness on the scale of separation between the contours.

We start with the former. In what follows we compute the contributions
to the correlation functions due to the pole term only. As discussed
above the cut contribution should not change any qualitative aspect of
our results but merely the overall coefficient by a number of order 1.
As before in order to compute the the two-point function for the
winding number we can show that only the integrals of the type $I_1$
survive, i.e., we have to compute
\begin{equation}
I_1 = \int dl_i dl_j  e^{i k.(\vec{ L}- \vec{ l})} \delta_{ij} =
l L \int_{-\pi}^{\pi} d\theta d\theta^\prime e^{i k_\|(L
\cos\theta^\prime -l \cos\theta)} \cos(\theta^\prime -\theta).
\end{equation}
This can be shown to yield
\begin{equation}
I_1 = 4 \pi^2 l L J_1 (k_\| L) J_1(k_\| l).
\end{equation}
We then see that in this case the correlation function is no longer,
necessarily, positive definite but rather depends on the interplay between the
radius of the two contours. The expression for the correlation function
then becomes
\begin{equation}
\langle N_\gamma N_{\gamma^\prime} \rangle = {e^2 \over 4
\pi^2} L l
\int k_\|
dk_\| d k_z {J_1(k_\| L)  J_1(k_\|l) \over \omega} n(\omega).
\end{equation}
Figure 5 shows the result of the numerical integration of (39), having
fixed the radius of one of the contours to be $l= {1\over T},{5\over
T}$ and ${10\over T}$ and $m=0.2T$.
It shows that the winding number is most correlated when the radius of
the two contours is comparable.
Moreover, when the difference between the radii of the two contours
becomes appreciable the correlation function exhibits the approximate
behavior
\begin{equation}
\langle N_\gamma N_{\gamma^\prime} \rangle \simeq {e^2 \over 4 \pi^2}
LT0.05 {e^{- 1.25 m(L-l)}
\over  \sqrt{L-l}}.
\end{equation}
Thus, for radial separations larger than essentially a few units of
the inverse mass,  the winding number
through the two contours becomes essentially uncorrelated showing that
it should correspond to different strings.  String, if
produced by a thermal fluctuation, would
therefore be expected to have a curvature radius smaller than that
scale, corresponding necessarily to loops of that form.

Another possible generalization of (25) corresponds to translating the
two contours, in the direction perpendicular to the plane they define,
relative to each other so
that they now define the two basis of a cylinder of height $h$, see fig.4.
In this case the form of the
original integral  still holds but the Fourier transform
corresponding to the 'z'-direction ceases to be trivial. We then obtain
\begin{equation}
\langle N_\gamma N_{\gamma^\prime}\rangle = {e^2 \over 4 \pi} L^2
\int_0^\infty k_\| d k_{\|} J_1^2( k_\| L)
\int_0^\infty d k_z e^{i k_z h} {n(\omega) \over \omega},
\end{equation}
\noindent where $h$ is the separation between the two contours along the
vertical axis. This  gives
\begin{equation}
\langle N_\gamma N_{\gamma^\prime}\rangle = 2 {e^2 \over 4 \pi} L^2
 \int_0^\infty k_\| d k_{\|} J_1^2( k_\| L)
\int_{0}^{\infty} d k_z \cos( k_z h) {n(\omega) \over \omega}.
\end{equation}
Figure 6 shows the result of these integrations for several values of
$L$ and $m=0.2T$.
Again an approximate analytical behavior can be found for (42).
The best fit on the mass and $h$ dependences yields
\begin{equation}
\langle N_\gamma N_{\gamma^\prime}\rangle \simeq {e^2 \over 4 \pi} 0.6 LT
e^{-1.25mh}
\end{equation}

Both correlation functions (39) and (42), above exhibit exponential
fall-offs with contour separation, determined by a correlation length
multiple of the inverse gauge field mass. This behavior strictly
holds for $m(L-l),mh\rangle\rangle1$. Together with the result
of the previous section about the magnitude of fluctuations, these
allow us to characterize the temperature for which winding number
fluctuations are relevant on a given length scale that can in turn be
compared to the string's proper length.
In is clear that, as the temperature drops,  only the
population of string loops smaller than $\xi= O( {1 \over m(T)})$
will be affected. Thus any long/infinite string emerging from the
phase transition will  be very little changed by fluctuations.
This is an important result as it states that the formation of long
string  is
essentially a classical process occurring at the level of the
Euler-Lagrange equations for the theory.

\section{Fluctuations in Winding Number in the presence of Strings}

In this section we investigate how a non-homogeneous
scalar modulus changes the fluctuations in winding number, as computed in
section 3. This is an obvious requirement of any description of
fluctuations taking place during a phase transition. Inhomogeneities
occur naturally in the dynamics of domain growth and coalescence,
corresponding to such interfaces as well as to topological
defects\footnote{In the early stages of this dynamics such a
distinction probably fails to apply.}.  As
the phase transition approaches completion we expect the domain
structure to coarsen and smooth away and the only inhomogeneities left
behind should correspond to topological defects. At all stages, thermal
fluctuations in winding number  will occur and a  freeze-out in
defect number will correspond to the instant in time when their role
is rendered negligible, in some well-specified way.

In what follows we will attempt to give a general treatment of
fluctuations in the presence of a weakly inhomogeneous scalar modulus
by systematically expanding the gauge field propagator in a small
perturbation around a homogeneous scalar field. We will later
specialize to the case where strings constitute such inhomogeneities.
As before we will only keep the pole term in the gauge field propagator.

Let us consider the case of a localized deviation from the otherwise
homogeneous background so that we can write the scalar modulus as
\begin{equation}
\varphi(x) = \eta - f(x)
\end{equation}
\noindent where $\eta$ is the background value of the scalar field
modulus and $f(x)$ the small perturbation.
Then we can expand the gauge propagator around the homogeneous
background field. We restrict ourselves to the relevant part of the
transverse propagator, for the sake of clarity. Formally, we obtain
\begin{eqnarray}
G_{\mu \nu}(x,y,M) &=& \langle x|{-\delta_{\mu \nu} \over \Box^2 +m^2
+  M^2(z)}|y\rangle \simeq \langle x|{-\delta_{\mu \nu} \over \Box^2 +m^2  }
\left[ 1 -  |z\rangle M^2(z)\langle z| {1 \over \Box^2 +m^2} + \right.
\nonumber \\
&+& \left.  {1\over 2}|z\rangle M^2(z)\langle z|{1 \over \Box^2
+m^2}|z^\prime\rangle
M^2(z^\prime)\langle z^\prime|{1 \over \Box^2 +m^2} \right. \nonumber \\
&+& \left.  {\rm higher\ order\ terms }  \right] |y\rangle ,
\end{eqnarray}
\noindent where $M^2(x) = e^2 f(x)\left[2 \eta -f(x)\right]$.
The first term in the expansion is just the gauge field propagator in
the presence of a homogeneous, static scalar field in the broken
phase. When used to compute winding number correlation functions it
will generate the results of the previous two sections, when taken in
thermal equilibrium. The first correction to that can be written as
\begin{equation}
G^{(1)}_{\mu \nu}(x,y)= \delta_{\mu \nu} \int d^4 p d^4 k d^4 z d^4 z^{\prime}
G(p) e^{i p(z-x)} M^2(z,z^\prime)G(k) e^{i k(y-z^\prime)},
\end{equation}
\noindent where $M^2(z,z^\prime)=M^2(z^\prime)\delta^4(z-z^\prime)$.
This leads to
\begin{equation}
G^{(1)}_{\mu \nu}(x,y)= \delta_{\mu \nu}\int d^4 p d^4 k
G(p) e^{-ipx} M^2(k -p)G(k) e^{i k y },
\end{equation}
\noindent where
\begin{equation}
M^2(k-p) = \int d^4 z e^{-iz(k-p)} M^2(z),
\end{equation}
\noindent is the usual Fourier transform of $M^2(z)$.
We see  that the presence of inhomogeneities can be accounted
for as the occurrence of sources, in the usual way. Their effect is
naturally to change
the 4-momentum of photons from an in-going state to an out-going one.
The order in the expansion (45) then corresponds to the number of such
momentum transfers.
To actually compute corrections to our previous results
we have to specify the functional form of $M^2(x)$.
As an illustration of the effect of inhomogeneities we take the
simplest case of associating $M^2$ with one
straight string at rest, placed in the center of the loop.
That is,
\begin{equation}
 M^2(x) = e^2 f(x)\left[ 2 \eta -f(x) \right] \simeq 2 e^2 \eta f(x),
\end{equation}
\noindent where
 we specialize the form of $f(x)$ to correspond the the field
profiles of a string at distances larger than its width
\cite{Bett}
\begin{equation}
f(r)= k_s K_0(m_S r) \simeq k_S \sqrt{1 \over 2 \pi m_S r} e^{-m_S r},
\end{equation}  where $K_0$ is the modified Bessel Function of order
zero and $k_S$ is a constant that for critically coupled strings can
be shown to be $k_S = \vert n \vert \eta$, with $n$ as the string's
winding number. Here we assume strings to be the usual static
solutions of the effective 3-dimensional theory, obtained after the
usual frequency sums at finite temperature. As such all mass scales
exhibit their usual temperature dependence.
The exponential regime  corresponds to the asymptotic
form of the Bessel function for large arguments. In particular we see
that, in this regime, the corrections to
the homogeneous background propagator will be exponentially suppressed
by the distance to the string.

It is straightforward to generalize the situation for one string to an
arbitrary number of strings, or other well localized sources of
inhomogeneity with a definite profile.
 We simply write the scalar modulus as
the superposition of all sources of inhomogeneity as
\begin{equation}
\varphi = \eta^{ 1 - N} \Pi_{i=1}^N (N-f_i) \simeq \eta - \sum_{i=1}^N
f_i + {\rm higher \ order \ terms \ in \ f}
\end{equation}
Let us then proceed to compute the change in fluctuations brought
about by a single string, to first order.
The form (50), corresponding to a single static cylindrically-symmetric
string placed in the center of our circular contours, actually allows us to
compute its Fourier transform analytically; namely
\begin{eqnarray}
M^2(k-p) &\simeq & \int d^4 x e^{-i(p-q)x} 2 m_e^2 K_0(m_s r)
\nonumber \\
&=& 4 \pi^2  m_e^2 { \delta(k_0 - p_0) \delta(k_z -p_z) \over (k -
p)_\|^2 + m_s^2},
\end{eqnarray}
\noindent which is just the potential, in momentum space, for a
massive cylindrically symmetric field, as it should be.
Using this form in (47) yields the correction to the winding number
correlation functions

\begin{eqnarray}
\langle N_\gamma^2\rangle^{(1)} &=&  {e^2 \over 8 \pi^3}
(L m_e )^2  \int  dp_\| dk_\| dk_z
{J_1(p_\| L)J_1(k_\|L)\over w_q w_k} \left({p_\|^2 +k_\|^2 +m_s^2
\over \sqrt{ (p_\|^2 + k_\|^2 +m_s^2)^2 -4p_\|^2k_\|^2}} -1 \right)
\nonumber \\
&\times &
\left[{1 \over w_p + w_k}\left( n(w_p) +n(w_k)\right) - {1\over w_k -
w_p}\left( n(w_k) - n(w_p) \right) \right],
\end{eqnarray}

\noindent where we have assumed that both the in-going and out-going
momenta display thermal distributions. This follows automatically
since a static source causes no energy shift.

The integral (53) has to be computed numerically. Its dependence on L,
together with that of the zeroth order
result of section 2 and their sum are depicted in fig. 7,
for ${m\over T}=0.5,{m_S \over T}=0.6$.
It shows that the effect  of this correction is only to change winding
number fluctuations significantly for small contour radius, smaller or
equal to the string's width. This can also be seen by fixing $L$ and $m_e$
and varying $m_s$. Indeed, the magnitude of the correction term
becomes larger the smaller
scalar masses we take, i.e., with greater string widths,  signaling the
breakdown of the weakly inhomogeneous scalar field expansion for the
gauge propagator.
In this regime the sum of the two terms
yields a negative quantity showing the clear breakdown of the expansion
(45) since the full quantity is positive definite. For large contour
radius the effect of the correction becomes increasingly more
negligible and the negative shift probably becomes a genuine feature.

Another relevant issue is to know how this correction term behaves
close to the critical temperature. This can be probed by assuming that
both masses vanish approximately as we  approach the critical
temperature. In practice we know this to be untrue for the gauge field
due to higher order effects, as discussed in section 3.
We can distinguish two cases, namely, when
the Yukawa gauge coupling is larger or smaller than the scalar
self-coupling. This differentiates between a type I and a type II
theory, respectively. Fig 8, shows the dependence of the zeroth
order term on the gauge mass as well as that of (54) for
$m_S=1.5m_e,m_e,0.5m_e$, for fixed $L=5/T$. We see that the magnitude
of the correction term grows with the gauge coupling, while still
remaining
 smaller than the zeroth
order term. This however does not hold for strong type I
theories for which we see that larger values of $L$ will be required
for the series expansion to hold. One important feature of the
correction term, that holds regardless of the hierarchy of the
couplings in the theory, is that it vanishes, at critical temperature,
with the scalar mass.
This should constitute a genuine behavior of all correction terms since
they are proportional to increasing powers of $M(x)$ . Thus as the
latter goes to zero so does $M(x)$ and we recover the usual
electromagnetic propagator.
This allows to conclude that the weakly inhomogeneous
background expansion can always be trusted  in a sufficiently small
neighborhood of the critical temperature and that the effect of the
presence of strings will manifest itself mostly at  intermediate
and low temperatures. These conclusions would radically change if we do not
adopt a thermal equilibrium mass scale (vanishing at the critical
temperature) in our defect
configurations. In
assuming this we hope to mimic an average defect configuration in
the presence of a thermal plasma and thereby  render the whole picture
self-consistent.

\section{Conclusions}

In this paper we have developed techniques to allow us to study the
role of  fluctuations in cosmic string formation in an Abelian
gauge theory.
Their application to the simplest case of quasi-homogeneous
scalar field backgrounds leads us to conclude that thermal
fluctuations in winding number
are strongest the closer the system is to  the critical temperature
and, further, that the relative winding number fluctuations
diverge at that temperature signaling the presence of no stable
defects, in these circumstances. This effect persists even in the
presence of inhomogeneities in the fields, such as strings themselves,
if thermal equilibrium is used throughout. As the temperature drops
the role of thermal fluctuations in changing winding number becomes
less and less important. The energy scale that characterizes an
approximate freeze-out in defect numbers is most naturally the inverse mean
domain size at the time of coalescence, which in turn is associated to
the average string interseparation. From the point of view of the
field theory its determination necessarily implies a more detailed
knowledge of the field dynamics during the transition than what is
currently known.

On developing generalizations of our initial winding number correlation
function we  also
conclude
that string configurations can only be changed by thermal fluctuations
on scales of the order of the thermal correlation length of the
fields. As such the population of small string loops can be drastically
modified but not that of long and infinite string, which is the most
relevant input for network numerical evolutions. A change in loop
population is only likely, for the present cosmic string network
implementations, to change the approach to the scaling regime and will
therefore bring no change to the details of late time structure
seeding in the canonical scenario.

An analogous winding number correlation function over the same contour
can be defined using the phase of the Higgs field
directly \cite{Riv}. Using the gauge sector instead  has great
advantages
in computational simplicity and  reliability as the former is
divergent as the symmetric phase is approached from below. This
leads to infinities and computability can only be maintained by
on adopting severe approximations.
Both  methods will necessarily measure fluctuations in
the gauge and scalar fields, respectively, that do not necessarily
correspond to strings. This is because, even though one of the fields
can have the
right configuration to generate a string, it will take the two at the
same point in space to
produce a Nielsen-Oleson vortex. One is then overcounting strings as
we will be computing the number of strings present at a given instant
together with that of fluctuations that can potentially produce a
string or decay away
\footnote{Other criteria exist to count the number of strings and their
fluctuations, namely those inspired in counting the density of zeros
of the scalar field \cite{GR}.  However, the same methods have yet
to be applied to gauge theories.}.

Our methods should have other applications beyond the GUT transitions.
The system considered above has obvious analogies with a superconductor
where the Ginzburg-Landau free energy is given precisely by a
non-relativistic analogue of the Abelian Higgs model. Our analysis
for the winding number correlation functions generalizes to
both models provided the mean field masses and self-energies get
properly translated. This has yet to be done, but would possibly constitute the
natural test
on the approach developed above as comparison with experiment  seems a feasible
task.
The next closest example of string configurations in a specific model
concerns those in the Electroweak Standard Model. Their study in the
simplest approach consists in treating the model as an effective
Abelian theory of a complex scalar field and the $Z$ gauge boson. Even
though the supplementary field excitations are known render them
unstable \cite{EWst} our  winding number correlation functions should find
applicability here, too.

The eventual applicability of this approach in genuinely
out-of-equilibrium scenarios hinges
upon a better knowledge of the corresponding fields and their two-point
correlation functions. A considerable amount of effort is
being devoted to this problem \cite{Boya, Us} and it is conceivable
that the full problem will be tractable in the near future.

\section*{Acknowledgements}

We would like to thank Luis J. Garay for numerous discussions and
help in the earliest stages of the numerical work. We would also like
to thank Tanmay Vachaspati for the discussions that lead to  section
3. The idea asking for string distributions in terms of conditional
probability statments is due to him.
L.M.A.B.'s research is supported by J.N.I.C.T. under contract BD 2243/92.
T.S.E. would like to thank the Royal Society for their support.

\clearpage

\section*{Figure Captions}

\vspace {.1in}

{\bf Figure 1} : The dependence of the winding number fluctuations on
the contour radius $L$ from the pole contribution,
for $m=0.2T,0.5T$ and $0.8T$.

\vspace {.1in}

{\bf Figure 2} : The dependence of the winding number fluctuations on
the gauge field mass from the pole contribution, for $L=T,5T$ and $15T$.

\vspace {.1in}

{\bf Figure 3} : A Schematic view of the two concentric contours and
of the strings that contribute to the winding number correlation
function between them.

\vspace {.1in}

{\bf Figure 4} : The same as in Figure 3 for the two coaxial circular contours.

\vspace {.1in}

{\bf Figure 5} : The dependence of the winding number
 correlation function between the two concentric contours on one of
the contour's radius $L$, for $m=0.2T$ and $l=T,5T$ and $10T$.

\vspace {.1in}

{\bf Figure 6} : The different curves exhibit the dependence of the
winding number correlation function between the two circular coaxial
contours on their separation $h$,for $m=0.2T$ and $L=5T,10T$ and $15T$.

\vspace {.1in}

{\bf Figure 7} :The magnitude of the zeroth order term (dotted line),
the absolute value of its first
correction and their sum as a function of contour radius and for
$m_S=0.4T$ and $m=0.2T$.

\vspace {.1in}

{\bf Figure 8} : The dependence of the zeroth order term in winding
number fluctuations (dotted line) and its first order correction
on the gauge field mass $M$,
for $L=5/T$ and in the cases $m_S=0.7m, m$ and $1.2m$. The mass scale
$M$ is in units of temperature.


\begin{thebibliography}{99}

\bibitem{Kib} T. W. B. Kibble,  J. Phys.  A {\bf 9}  1387 (1976).

\bibitem{Book} For a comprehensive review see A. Vilenkin and
E. P. S. Shellard, {\it Cosmic Strings and
other Topological Defects} (Cambridge: Cambridge University Press, 1994).

\bibitem{Helium} P. C. Hendy, N. S. Lawson, R. A. M. Lee, P. V. E.
McLintock and C. D. H. Williams, Nature {\bf 368} 315 (1994).

\bibitem{liqcrys} I. Chuang, R. Durrer, N. Turok and B. Yurke, Science
{\bf 251} 1336 (1991); M.J. Bowick, L. Chander, E. A. Chander, E. A. Schiff
and A. M. Srivastava, Science {\bf 263} 943 (1994).


\bibitem{Struc}Ya. B. Zeldovich, Mon. Not. Astron. Soc. {\bf 192} 663
(1980);  A. Vilenkin, Phys. Rev. Lett.  {\bf 46} 1169 (1981);
 {\bf 46} 1496(E) (1981).

\bibitem{VV} T. Vachaspati and A. Vilenkin, Phys. Rev. D
{\bf 31} 3052 (1985).

\bibitem{KibVil} T.W.B. Kibble and A. Vilenkin, Imperial College and Tufts
University preprint  Imperial-TP-94-95-9A, TUTP-95-1.

\bibitem{Brand} R. H. Brandenberger and A.-C. Davis, Phys. Lett. B {\bf 332}
 305 (1994).

\bibitem{Riv} R. J. Rivers, in  {\it Electroweak Physics and the Early
Universe},  Ed. by J. C. Romao and F. Freire, Plenum Press, New York
(1994) .

\bibitem{NO} H. B. Nielsen and P. Oleson, Nucl. Phys. B
{\bf 61} 45 (1973).

\bibitem{Sriv} S. Rudaz and A. Srivastava, Mod. Phys. Lett. A {\bf 8}
1143 (1993).

\bibitem{Boya} D. Boyanovski, H. J. de Vega, R. Holman, D.S. Lee
and A. Singh, Phys. Rev. D {\bf 51} 4419 (1995).

\bibitem{BP} Braaten and Pisarski, Nucl. Phys. B {\bf 337} 569 (1990).

\bibitem{Braaten} E. Braaten and T.C.Yuan, Phys. Rev. Lett. {\bf 66}
2183 (1991).


\bibitem{Cop} E. Copeland, G. Cheetham, T. S. Evans and R. J. Rivers,
Phys. Rev. D {\bf 47}  5316 (1993).

\bibitem{Mazenko} More  can be done in spin models  and other condensed
matter systems. See e.g. G. F. Mazenko, W. G. Unruh and R. M. Wald,
Phys. Rev D {\bf 31} 273 (1985).

\bibitem{Bett} L. M. A. Bettencourt and R. J. Rivers, Phys. Rev. D
{\bf 51} 1842 (1995).

\bibitem{GR} A. J. Gill and R. J. Rivers, Imperial College preprint
IMPERIAL-TP-93-94-55.

\bibitem{EWst} M. James, L. Perivolaropoulos and T. Vachaspati Nucl.
Phys. B {\bf 395} 534 (1993).

\bibitem{ Us} N. D. Antunes, L. M. A. Bettencourt and G. Karra, in progress.

\end{thebibliography}
\end{document}